\documentstyle[preprint,aps,eqsecnum]{revtex}
\begin{document}
\draft
\title{Relativistic Two-Body Processes \\
in Axial-Charge Transitions}
\author{E.D. Izquierdo}
\address{Departamento de F\'{\i}sica, Facultad de Ciencias Exactas y
Naturales, \\
Universidad de Buenos Aires. Pabell\'on 1, Ciudad
Universitaria,
(1428) Buenos Aires, Argentina}
\author{G. Barenboim and  A.O. Gattone}
\address{Departamento de F\'{\i}sica, Comisi\'on Nacional de Energ\'{\i}a
At\'omica \\ Av. del Libertador 8250, 1429 Buenos Aires, Argentina}
\date{December 19, 1994}
\maketitle
\begin{abstract}
We study the contribution of two-body meson-exchange processes to
axial charge transitions for nuclei in the lead, tin and oxygen
regions.  We conduct calculations in the Dirac-Hartree (the Walecka
model) and the relativistic Hartree (where the full one-nucleon-loop
effects are included) approximations.  We present results indicating
that one- and two-body processes enhance the matrix elements of the
axial-charge operator by some (100$\pm$20) \% in all three regions
studied.  This agrees well with the fit of eighteen first-forbidden
$\beta$-decay transitions conducted by Warburton in the lead region.
We also discuss some sensitivities present in the calculation.
\end{abstract}



\section{Introduction}

The unexpected enhancement found in processes driven by the axial
charge, namely the cases of first-forbidden beta decays in heavy
nuclei~\cite{warburton} and the production of pions near threshold in
$pp\rightarrow pp\pi^0$ reactions~\cite{meyer}, has given a new
impetus to the study of meson-exchange currents (MEC) in axial charge
transitions.  The findings are really remarkable.  In the first
process for example ---already known to be strongly renormalized by
the presence of MEC--- the results found by Warburton~\cite{warburton}
exceeded the anticipated theoretical estimate by a factor of two.  In
the second ---a new result in a region not available to experiment
before--- the data overestimate the theoretical predictions (only the
one-body contribution) by approximately a factor of five.  Despite
this sizable difference between theory and experiment the first result
is acknowledged as the most dramatic since the process of
first-forbidden $\beta$-decay was thought to be (at least to a large
extent) under theoretical control~\cite{towner1}.

The subject of axial charge transitions becomes more relevant from the
viewpoint of relativistic theories of the nucleus.  Most of the
explanations advanced so far to account for these enhancements
(refs.~\cite{riska,towner} in the first case and~\cite{chuck} in the
second) rely on the presence of scalar heavy-meson exchange between
two nucleons.  Diagrams containing a nucleon-antinucleon pair
originating (or decaying) into a scalar meson of $m_s\approx$ 500 MeV
using couplings from either the Bonn or Paris potentials seem to
take care of the discrepancy between the theoretical prediction and
the experimental result.  These large Lorentz scalar potentials have
long been a distinctive feature of relativistic calculations whose
explicit need, however, had not shown up in the non-relativistic
phenomenology of nuclear processes before.  The calculations alluded
to above (non relativistic) incorporate the scalar degree of freedom
formally via meson-exchange currents.

The connection between non-relativistic pair-exchange currents and
relativistic mean-field theories for isovector currents was studied by
Delorme and Towner~\cite{delTow} in their paper of 1987 (in the case
of isoscalar currents there were previous calculations by Arima {\em
et al.}~\cite{arima} and Blunden~\cite{blunden}).  Pair graphs are
shown in Fig.\ref{fig1} (diagrams (a) and (b)).  The relativistic
Hartree contribution is obtained by closing the line on nucleon 2 and
is shown in diagrams (c) and (d) (diagram (d) is usually recognized as
the ``backflow" correction).  The difference between both approaches
is that whereas in the relativistic case the propagation of an
antinucleon is included in the one-body Green's function, in the
non-relativistic case it can only be incorporated as an exchange
current.  Delorme and Towner found in their calculations large
enhancements due to the scalar-exchange terms.  Their conclusion at
the time ---for axial-charge transitions--- was that such enhancement
(and along with it the relativistic mean-field theory) was hardly
consistent with experiment.  Quite oppositely, they noticed that the
conventional theory of meson exchange with short range cut-offs gave
reasonable agreements with the data (ignoring, of course, the
contribution of pair diagrams).  The more recent work of Warburton on
the one hand and the results of Meyer {\em et al.} on the other,
prompted a reanalysis of the existing calculations under the light of
the new data.  At the present time it is generally held that scalar
heavy-meson exchanges can account for the discrepancies between the
predictions and the actual data~\cite{warbtow}.

This situation opened up the ground for further exploration of the
predictions of relativistic models of nuclear structure to
axial-charge processes.  Since the physical content of the two
descriptions (non-relativistic vs. relativistic) is different from the
outset, the individual contribution of one- and two-body diagrams is
not anticipated to be similar.  We already mentioned that the one-body
relativistic calculation includes the non-relativistic two-body pair
graphs.  Conversely, the two-body contribution in the relativistic
calculation sums diagrams (like the one shown in Fig.\ref{fig2}) that
are not included in the traditional meson-exchange calculations.
Though, {\em a priori}, contributions of this type are small a
comparision between both approaches will only be meaningful once full
calculations of one and two-body processes are conducted within each
framework.

With this in mind we study in this paper the contribution of two-body
processes within the relativistic approach.  We concentrate
exclusively on first-forbidden beta decays in nuclei (oxygen, tin and
lead) where the relativistic mean-field can be used reliably.
Regardless of the model the symmetries must be upheld.  In particular
chiral symmetry, namely the use of soft-pion theorems and current
algebra, imposes a strong renormalization of the matrix elements due
to pion-exchange processes (particularly through the
$\pi\rho$-exchange diagram.)  This mechanism needs to be
incorporated explicitly into relativistic calculations since, at the
mean-field level, the pion is absent due to parity arguments.  It is
clear that a relativistic description of axial-charge transitions
in nuclei requires to go beyond the mean-field.  A first approximation
to the problem calls for a mean-field calculation (impulse
approximation) to describe the one-body processes and additionally a
MEC calculation taking care of the ``backflow" and pion-exchange
processes.  We notice that the results of this programme are not
expected to be free from the usual ambiguities of pion couplings in
relativistic theories (for a review of the current status see
ref.~\cite{serot2}).

Recently, within the framework of a relativistic model (quantum
hadrodynamics II) we analyzed axial transitions~\cite{izquierdo} in
heavy nuclei using the scheme outlined above.  We obtained first the
one-body contribution given by graphs (c) in Fig.~\ref{fig1}; next,
the RPA-type vertex corrections (graphs (d) in the same
figure)~\cite{izquierdo1} and, finally we investigated the
contribution of the pion-exchange currents~\cite{barenboim} (with no
exchange diagrams.)  Inspection of the results arrived at in this
study showed that:  First, despite some ambiguities related to the
strength of the $\rho$N coupling constant, the sum of one- and
two-body diagrams added up to an enhancement which was comparable to
that obtained in non-relativistic calculations.  Second, the
theoretical calculations could only be brought into agreement with the
data when pseudovector $\pi$N coupling was used (otherwise the results
were anomalously large).  Third, two-body pion-exchange processes were
indispensable to understand the axial-charge data.  These conclusions
were based on calculations conducted on two particular
transitions, one in Tl and the other in Pb, and were intended to give
an idea of what should be expected for other transitions in this mass
region.  In this paper we extend and deepen the analysis of the
previous works.  On the one hand, we give a complete description of
the two-body pion-exchange contribution to the axial charge from the
point of view of relativistic hadron theories.  On the other, we
extend the results of~\cite{barenboim} and present results of MEC
contribution to axial charge matrix elements for tin and oxygen.  The
idea is to move to lighter nuclei to study mass dependence, and the
influence of the scalar field in this contribution.

Besides this introduction the paper contains two other sections.
Section~\ref{MECme}, divided into two subsections, deals with the
calculation of the matrix elements of the axial charge.  In the first
subsection we give some details about the hadronic model we employ and
the way the axial two-body current is derived.  In the second we describe
at some length the scheme to carry out the actual calculation of the matrix
elements of the current between the nuclear states.  The last section
contains a discussion of the results and of the potential ambiguities
involved in the calculation. Finally we present some conclusions.

\section{Meson-exchange-current matrix elements of the axial charge}
\label{MECme}
\subsection{Hadron model and currents}

At tree level the two-body isovector axial vector current is given by
the sum of the four diagrams shown in Fig.~\ref{fig3}.  In the figure
we have used the symbol $\varepsilon$ to denote the field operator
mediating the weak current.  Diagrams $a)$ and $b)$ turn out to be the
most important since they involve pion exchanges which generate the
longest-range exchange currents. The contribution of the other two
diagrams (involving the $\Delta$) is known to be small~\cite{nozawa}.

The two-body current from diagrams $a)$ and $b)$ can be calculated
using standard Feynman rules.  As stated meson and baryon propagators
derived from some baryonic Lagrangian are needed. Throughout the paper
we use the {\em quantum hadrodynamical model II}\ (QHD--II) of Serot
and Walecka~\cite{serot}. It includes the nucleon fields, two isoscalar
meson fields --the scalar $\sigma$ and the vector $\omega$-- and two
charged isovector meson fields --the pseudoscalar $\pi$ and the vector
$\rho$.  The full Lagrangian expression is,
\begin{eqnarray}
{\cal L}&=&\bar{\psi}\left[i\gamma_{\mu}D^{\mu}-(m_N-g_\sigma \sigma)-
g_{\omega}\gamma_\mu\omega^{\mu}-ig_\pi\gamma_5\vec{\tau}\cdot\vec{\pi}
\right]\psi \nonumber \\
& & +g_{\rho}(\partial^\mu\vec{\pi}\times\vec{\pi})\cdot\vec{\rho}_{\mu}
+\frac{1}{2}g_{\rho}^2(\vec{\pi}\times\vec{\rho}_{\mu})\cdot
(\vec{\pi}\times\vec{\rho}_{\mu})+ \frac{1}{2}(\partial_\mu\sigma
\partial^\mu\sigma -m_\sigma^2\sigma^2) \label{Lagrangian} \\
& & -\frac{1}{4}F_{\mu\nu}F^{\mu\nu}+ \frac{1}{2}m_\omega^2
\omega_\mu\omega^\mu - \frac{1}{4}\vec{B}_{\mu\nu}\vec{B}^{\mu\nu}
+ \frac{1}{2}m_\rho^2\vec{\rho}_\mu\cdot\vec{\rho}^\mu\nonumber \\
& & + \frac{1}{2}(\partial_\mu\vec{\pi}\cdot\partial^\mu\vec{\pi}
-m_\pi^2\vec{\pi}\cdot\vec{\pi})+\frac{1}{2}g_{\sigma\pi}m_{\sigma}
\sigma\vec{\pi}\cdot\vec{\pi} \nonumber
\end{eqnarray}
In Eq.(\ref{Lagrangian}) the neutral and charged vector meson field strengths
are defined by,
\begin{equation}
F^{\mu\nu}\equiv\partial^\mu\omega^\nu-\partial^\nu\omega^\mu
\end{equation}
\noindent
and
\begin{equation}
\vec{B}_{\mu\nu}\equiv\partial_\mu\vec{\rho}_\nu-\partial_\nu\vec{\rho}_\mu
-g_\rho(\vec{\rho}_\mu\times\vec{\rho}_\nu)
\end{equation}
respectively. We have also made use of the covariant derivative
defined as
\begin{equation}
D_{\mu}\equiv\partial_{\mu}+i\frac{1}{2}g_{\rho}\vec{\tau}
\cdot\vec{\rho}_{\mu}.
\end{equation}
There is a Higgs--meson sector associated with the $\rho$--meson field
which has been suppressed by taking the scalar Higgs mass to be very
large.

The $\pi N$ coupling in the QHD-II Lagrangian is pseudoscalar.  To
account for the small pion scattering length in $\pi N$ interactions a
$\sigma\pi\pi$ coupling term is added.  As we will see below
pseudoscalar coupling with the standard parameters gives very large
predictions for the contribution of one- and two-body processes.  As
an alternative to the actual calculation we have considered
substituting pseudovector for both pseudoscalar $\pi N$ and
$\sigma\pi\pi$ couplings.  We emphasize that despite the different
predictions both couplings show the same correct behaviour in the
low-energy limit (in the two-body case) and both satisfy the
constraint of partial conservation of the axial current (PCAC).

The strong vertices $\pi NN$, $\rho N N$ and $\rho \pi \pi$ as derived
from the Lagrangian of QHD-II are shown in Fig.  \ref{fig4}.  The {\em
weak} vertices $\pi \varepsilon$ y $\pi \rho \varepsilon$ (see
Fig.~\ref{fig5}) cannot obtain directly from the QHD-II Lagrangian
since it does not contain the weak interaction.  The structure of the
$\pi\varepsilon$ vertex is obtained from symmetry arguments in pion
decay where,
\begin{eqnarray}
\Gamma_{\pi \epsilon}^{\mu\, a b}\, =\,
\langle\, \mbox{\em vac}\, |\, J^{A, \mu a}\, |\, \pi^b(q)\,
\rangle\, =\, i f_\pi q^\mu \delta^{a b}.
\end{eqnarray}

\noindent
The value of the constant, $f_\pi = 0.93$ GeV, is taken from
experiment~\cite{Walecka Muon}.  For the $\pi \rho \epsilon$ vertex a
similar argument leads to a vertex of the form,
\begin{eqnarray}
\Gamma_{\pi \rho \epsilon}^{\mu \nu\, a b c}\, =\,
\langle\, \pi^a(k_1)\, \rho^b(k_2)\, |\, J^{A, \mu c}\, |\,
\mbox{\em vacuum} \rangle\,
=\, \epsilon_{a b c}\, T^{\mu \nu}
\end{eqnarray}

\noindent
where now the structure and content of the $ T^{\mu \nu}$ tensor
cannot be determined from Lorentz covariance only.  However we will
see below that the requirement of $PCAC$ on the two-body axial current
suffices to define it.

The propagators of the $\pi$ and $\rho$ fields appearing in QHD-II are
of the Klein-Gordon type.  With them and the vertices just defined we
can obtain expressions for the contributions to the two-body axial
vector current density following Feynman prescriptions for diagrams
$a)$ and $b)$ in Fig.~\ref{fig3}.  For the case of pseudovector $\pi
N$ coupling they read,
\begin{eqnarray}
J^{A, \mu\, c}_{\rho \pi} &=& - i \frac{g_\pi g_\rho}{4 M}\,
\left[ \tau_1 \times \tau_2 \right]^c\,
\bar{\psi}(p'_1)\, \gamma^5\, \gamma_\lambda\, k_1^\lambda\, \psi(p_1)\,
\nonumber \\
& & \times \frac{1}{k_1^2 - m_\pi^2}\, T^{\mu \nu}\, \frac{1}{k_2^2 -
m_\rho^2}\,
\bar{\psi}(p'_2)\, \gamma_\nu\, \psi(p_2) \label{JA_ropi}
\end{eqnarray}
for diagram $a)$ and,

\begin{eqnarray}
J^{A, \mu\, c}_{\rho \pi \pi} &=& - i \frac{f_\pi g_\pi g_\rho^2}{4 M}\,
\left[ \tau_1 \times \tau_2 \right]^c\,
\bar{\psi}(p'_1)\, \gamma^5\, \gamma_\lambda\, k_1^\lambda\, \psi(p_1)\,
\frac{1}{k_1^2 - m_\pi^2} \nonumber \\ \nonumber \\
& & \times q^\mu\, \left( q + k_1 \right)^\nu\, \frac{1}{q^2 -
m_\pi^2}\, \frac{1}{k_2^2 - m_\rho^2}\,
\bar{\psi}(p'_2)\, \gamma_\nu\, \psi(p_2). \label{JA_ropipi}
\end{eqnarray}
for diagram $b)$.

\noindent
The sum of (\ref{JA_ropi}) and (\ref{JA_ropipi}) gives the full
two-body (2B) isovector axial-vector current,
\begin{eqnarray}
J^{A, \mu\, c}_{2B} &=& - i \frac{g_\pi g_\rho}{4 M}\,
\left[ \tau_1 \times \tau_2 \right]^c\,
\bar{\psi}(p'_1)\, \gamma^5\, \gamma_\lambda\, k_1^\lambda\, \psi(p_1)\,
\frac{1}{k_1^2 - m_\pi^2}\, \frac{1}{k_2^2 - m_\rho^2}\,
\nonumber \\
& &
\times\left\{ T^{\mu \nu}\, +\, g_\rho f_\pi \frac{q^\mu}{q^2 - m_\pi^2}
\left( q + k_1 \right)^\nu\, \right\}\,
\bar{\psi}(p'_2)\, \gamma_\nu\, \psi(p_2) \label{JA_2C}
\end{eqnarray}

\noindent
The quantity between brackets behaves as the effective $\pi \rho
\varepsilon$ vertex with the tensor $T^{\mu\nu}$ still an unknown.
This last difficulty can be overcome after enforcing the requirement
of PCAC on two nucleons. This is equivalent to demand
that the four-divergence of the full two-body current be equal to,

\begin{eqnarray}
q_\mu\, J^{A, \mu\, c} &=& i\, f_\pi\, \frac{m_\pi^2}{q^2 - m_\pi^2}\,
M^c(2B)  \label{PCAC_M}
\end{eqnarray}

\noindent
In (\ref{PCAC_M}),
$M^c(2B)$ is the $\pi$-absorption amplitude on two nucleons
which obtains from the
$\rho \pi \pi$ diagram (diagram $b)$ in Fig.~\ref{fig3}) by removing
the external
$\varepsilon^{5 \mu\, a}$ line. Thus from the figure,
\begin{eqnarray}
M^c(2B) &=& - \frac{g_\pi g_\rho^2}{4 M}\,
\left[ \tau_1 \times \tau_2 \right]^c\,
\bar{\psi}(p'_1)\, \gamma^5\, \gamma_\lambda\, k_1^\lambda\, \psi(p_1)
\nonumber \\ \nonumber \\
& & \times \left( q + k_1 \right)^\nu\, \frac{1}{k_1^2 - m_\pi^2}\,
\frac{1}{k_2^2 - m_\rho^2}\,
\bar{\psi}(p'_2)\, \gamma_\nu\, \psi(p_2). \label{M(2C)}
\end{eqnarray}

\noindent
Substitution of (\ref{M(2C)}) into the right hand side of
(\ref{PCAC_M}) gives for the $T^{\mu \nu}$ tensor the expression,
\begin{eqnarray}
T^{\mu \nu} &=& - f_\pi\, g_\rho\,
\left[\, g^{\mu \nu}\, +\, \frac{q^\mu k_1^\nu}{q^2 - m_\pi^2} \right]
\end{eqnarray}
\noindent
thus fixing the strength at the $\rho \pi \varepsilon$ vertex.  It is
important to notice that in the soft--pion limit ($k_1 \rightarrow
0$), one obtains $T^{\mu \nu} = - f_\pi\, g_\rho\, g^{\mu \nu}$.  This
is the same strength obtained by taking the soft--pion limit in
hard--pion models (see for example, Ivanov and Truhlik\cite{ivanov}.)

Substituting $T^{\mu \nu}$ into the expression for the axial current
and using the relation $f_\pi g_\pi = g_A M$ (Goldberger--Treiman),
\begin{eqnarray}
J^{A, \mu\, c}_{2B} &=& - i \frac{g_A g_\rho^2}{4}\,
\left[ \tau_1 \times \tau_2 \right]^c\,
\bar{\psi}(p'_1)\, \gamma^5\, \gamma_\lambda\, k_1^\lambda\, \psi(p_1)\,
\frac{1}{k_1^2 - m_\pi^2}\, \frac{1}{k_2^2 - m_\rho^2}\, \nonumber \\
& & \times
\left\{ - g^{\mu \nu}\, +\, \frac{q^\mu q^\nu}{q^2 - m_\pi^2} \right\}\,
\bar{\psi}(p'_2)\, \gamma_\nu\, \psi(p_2). \label{JA_2Cfin}
\end{eqnarray}

\noindent
which is our final expression for the full two-body isovector
axial-vector current. Inspection of the quantity between curly
brackets shows that $J^{A, \mu\,c}_{2B}$ satisfies $PCAC$ since,
\begin{eqnarray}
q_\mu\, J^{A, \mu\, c}_{2B} &\propto&
q_\mu\,
\left\{ - g^{\mu \nu}\, +\, \frac{q^\mu q^\nu}{q^2 - m_\pi^2} \right\}\,
=\, q^\nu\, \frac{m_\pi^2}{q^2 - m_\pi^2}\,\,
-\!\!\!\!-\!\!\!\!-\!\!\!\!-\!\!\!\!\longrightarrow_{\!\!\!\!\!\!\!\!
\!\!\!\!\!\!\!\!\!\!\!\!\!\!\!_{_{\small{m_\pi\, \rightarrow\, 0}}}}
\,\,\,\,\,\, 0
\label{PCAC_fin}
\end{eqnarray}
\noindent
Thus, it is clear that the two-body axial current derived from the
Lagrangian (\ref{Lagrangian}) satisfies PCAC and has the correct
behaviour in the soft-pion limit.

\subsection{Multipole decomposition}
\label{md}

Having studied the structure of the relativistic two-body current we
concentrate next on the nuclear structure.  Individual transitions in
nuclei are more conveniently studied by performing a multipole
decomposition of the current matrix elements.  For axial-charge
mediated processes selection rules~\cite{izquierdo} limit the
contributing operators to the axial-Coulomb (or charge) operator, and
the polar-vector longitudinal and transverse-electric operators.  The
last two affect only minimally the total matrix element and therefore,
though they will be taken into account in the actual calculations and
displayed in the results, for the purpose of this section we shall
work exclusively with the first one.  The axial-Coulomb operator is
written in the form,
\begin{equation}
\hat{\sf C}_{JM}^{2B}(q)\equiv
\int d^{3}x  M_J^M(q{\vec{x}}) J_0^{2B}(\vec{x}) \label{coul}
\end{equation}
with
\begin{equation}
M_J^M(q{\vec{x}})=j_J(qx) Y_J^M(\hat{x}).
\end{equation}

The reduced two-body matrix element of any multipole operator may be
decomposed as a sum of products of a matrix element of the operator
between two-particle states times a two-body density matrix
element which accounts for the complexity of the many-body problem.
Thus, for the axial Coulomb operator we shall write,
\begin{equation}
\langle J_f||\hat{\sf C}_{J}^{(2B)}(q)||J_i\rangle=
\sum_{a,b,j_1}\sum_{a',b',j_2}<[a',b']j_2||
\hat{\sf C}_{J}^{(2B)}(q)||[a,b]j_1>
\Psi^{(i,f)}_{J}\left ([a', b']j_2,[a,b]j_1\right) \label{c0}
\end{equation}
where the sums run over all initial (final) two-particle states
coupled to angular momentum $j_1$ ($j_2$). The two-body density
matrix element is written in terms of single particle creation
operators, $c_\alpha^\dagger$, in the form
\begin{eqnarray}
\Psi^{(i,f)}_{J}\left
([a',b']j_1,[a,b]j_2)\right)& = & \nonumber \\
& & \frac{1}{[J]}<J_i||\left[\left[ c^\dagger_{a'}\otimes
c^\dagger_{b'}\right]_{j_2}\otimes \left[\tilde{c}_a\otimes
\tilde{c}_b\right]_{j_1}\right]_J||J_f>
\end{eqnarray}
with $\tilde{c}_a=(-1)^{j_a+m_a}c_{-a}$.  (We refer the reader to
Sec.IV.E in Donnelly and Sick's review article~\cite{donnelly} for
the explicit expressions of the 2B-density matrix elements in the case
of one particle or one hole outside a closed shell which are those of
interest for this work.)

Equation ~(\ref{c0}) requires the calculation of the reduced matrix
element of the axial Coulomb operator between two-particle states,
\begin{equation}
|[a',b']_{j_1}^{m_1}\rangle= \sum_{m_a,m_b} \langle j_am_aj_bm_b|
j_1m_1\rangle |j_am_a\rangle|j_bm_b\rangle
\end{equation}
where
\begin{equation}
\langle\vec{x}|j_am_a\rangle=\Psi_{nk}^\mu(x) =\left(\begin{array}{c}
                               G_{nk}(x) {\cal Y}_k^\mu(\Omega)\\
                                                               \\
                              iF_{nk}(x) {\cal Y}_{-k}^\mu(\Omega)
                               \end{array}
                                 \right)
\exp{\{-iE_{n_al_a}t\}}
\end{equation}
with
\begin{equation}
{\cal Y}_k^\mu = \sum_{m,m_s} <l m \frac{1}{2} m_s | j \mu> Y_l^m
    \chi_{\frac{1}{2}}^{m_s}\nonumber\end{equation}
and
\begin{equation}
k_\alpha=(-1)^{j_a+l_a+\frac{1}{2}}\left(j_a+\frac{1}{2}\right)
\end{equation}
Since single particle states are obtained in co-ordinate space we
transform the current accordingly.

\subsubsection{Coordinate space two-body matrix elements}

Our starting point will be the total 2B current of Eq.(\ref{JA_2C})
(where, for convenience, all but the Lorentz indeces have been
dropped.) In the Heisenberg representation we write the
current in co-ordinate space in the following form,
\begin{eqnarray}
\hat{J}^\mu_a(x)&=&\frac{1}{4} g_A g_\rho^2 \epsilon_{abc}
\left\{- g^\mu_\nu\, -\, \frac{\partial^\mu \partial_\nu}{q^2 -
m_\pi^2} \right\}\nonumber \\
& & \times \int d^4x_1\;\Delta^0_\pi(x-x_1)
\;\partial^1_\alpha\;\left[\bar\psi(x_1)
\gamma^\alpha\gamma^5\tau^b\psi(x_1)\right] \nonumber \\
& & \times \int d^4x_2\;\Delta^0_\rho(x-x_2)\;
\left[-g^{\nu \eta}\right]\left[\bar\psi(x_2)
\gamma_\eta\tau^c\psi(x_2)\right]  \label{jcoor}
\end{eqnarray}
\noindent
where
\begin{equation}
\Delta^0_{\pi,\rho}= -\int \frac{d\kappa^0}{2\pi} e^{-i\kappa^0(t-t')}
\frac{e^{-M_{\pi,\rho}^{\kappa^0}\mid\vec{x}-\vec{x}'\mid}}{4\pi
\mid\vec{x} - \vec{x}'\mid}
\end{equation}
and $M_{\pi,\rho}^{\kappa^0}=+\sqrt{m_{\pi,\rho}^2-(\kappa^0)^2}$.
For the $\rho NN$ vertex we have worked in the t'Hooft-Feynman gauge.
We have kept a soft q-dependence in the second term of the
$\pi\rho\varepsilon$ vertex in (\ref{jcoor}) that will be dropped in
actual calculations. For future use we expand
the meson propagators into multipoles using,

\begin{equation}
\frac{e^{-m\mid \vec{x}-\vec{x}'\mid}}{4\pi\mid
\vec{x}-\vec{x}'\mid}=-m \sum_{l,m_l}^\infty
j_l(i m {r'}_<) h^{(1)}_l(i m {r'}_>)
Y_{l,m_l}^*(\Omega) Y_{l,m_l}(\Omega') \label{mp}
\end{equation}
where $j_l(z)$, and $h^{(1)}_l(z)$
are spherical Bessel functions, and ${r'}_<({r'}_>)$ denotes
the lesser (greater) of $\mid\vec{x}\mid$ and $\mid\vec{x}'\mid$.

To calculate the reduced matrix element $
<[a',b']j_2||\hat{\sf C}_{J}^{(2B)}(q)||[a,b]j_1>$  we
substitute the expansion (\ref{mp}) into (\ref{jcoor}) perform
the time and angular integrations and contract the Heisenberg
operator between the appropiate two-particle states. For convenience
we divide the total calculation into two separate contributions
stemming from each of the two terms in the $\pi\rho\varepsilon$
vertex. For the first term proportional to the time component of the
metric tensor (labelled A) we arrive at an expression of the form,
\begin{eqnarray}
& & <[a',b']j_2||\hat{\sf C}_{J}^{A}(q)||[a,b]j_1>=
\frac{i}{4\sqrt{\pi}}g_A g_\rho^2 \:\xi_\pm\:  \sqrt{2J+1} (-1)^{J+1}\; \int
dr\:  r^2\:j_J(qr) \nonumber \\
& & \times \int dr_1\: r_1^2\: M_\pi^{\omega_a} \sum_{l_1=0}^\infty\:
\sqrt{2l_1+1}\: j_{l_1}(iM_\pi^{\omega_a} r^1_<)
h_{l_1}(iM_\pi^{\omega_a} r^1_>)
I_{l_1}(k_{a'},-k_a) \nonumber \\
& &\times \left\{ \phantom{\frac{\stackrel{\leftrightarrow}{\partial}}
{\partial r_1}}\!\!\!\!\! (E_{a'}-E_a)[G_{a'}(r_1)F_a(r_1)-
F_{a'}(r_1)G_a(r_1)]\right. \nonumber \\
& &\;\;\;\;+\left.\left[G_{a'}(r_1)\left(\frac{\stackrel{\leftrightarrow}
{\partial}}{\partial r_1} + \frac{2}{r_1}\right)G_a(r_1) +
F_{a'}(r_1)\left(\frac{\stackrel{\leftrightarrow}{\partial}}
{\partial r_1} + \frac{2}{r_1}\right)F_a(r_1)\right]\right\}
\nonumber \\
& & \times \int dr_2\:  r_2^2 M_\rho^{\omega_b} \sum_{l_2=0}^\infty
\sqrt{2l_2+1}\: j_{l_2}(iM_\rho^{\omega_b}{r^2}_<)\:
h_{l_2}(iM_\rho^{\omega_b}{r^2}_>)\;
I_{l_2}(k_{b'},k_b) \nonumber \\
& & \times [G_{b'}(r_2)G_b(r_2)+F_{b'}(r_2)F_b(r_2)]
\left( \begin{array}{ccc}
         l_1 & l_2 & J \\
         0   & 0   & 0 \\
\end{array} \right)
\left\{ \begin{array}{ccc}
        j_{a'} & j_{b'} & j_2 \\
        j_a    & j_b    & j_1 \\
        l_1    & l_2    & J   \\
\end{array} \right\} \label{C(A)}
\end{eqnarray}
\noindent
where $M_{\pi,\rho}^{\omega_{a,b}}=
\sqrt{m_{\pi,\rho}^2-(E_{a,b}-E_{a',b'})^2}$, and
\begin{equation}
I_L(k',k)=\frac{(-1)^{j'+1/2}}{4\sqrt{\pi}}\sqrt{(2j+1)(2j'+1)(2L+1)}
\left[1+(-1)^{l+l'+L}\right]
\left( \begin{array}{ccc}
       j'          & L & j            \\
       \frac{1}{2} & 0 & -\frac{1}{2} \\
       \end{array} \right)
\end{equation}
\noindent
The symbol $\xi_\pm$ in (\ref{C(A)}) denotes the matrix element of
the charge-raising (or lowering) isospin current between
two-particle states,
\begin{eqnarray}
\xi_\pm &=&
<a'b'|\frac{1}{2 i}\left[\vec{\tau}_1\times\vec{\tau}_2\right]_\pm|a b>\,
=\, \pm <a' b'|(\tau_1^3 \tau_2^{\pm} - \tau_1^{\pm} \tau_2^3 )|a b>
\nonumber \\
&= & \pm \left[(\delta_{a'\,p}\, \delta_{a\, p}\, -\,
\delta_{a'\,n}\,\delta_{a\, n})\,
\delta_{b'\, ^p_n}\, \delta_{b\, ^n_p}\, -\,
\delta_{a'\, ^p_n}\, \delta_{a\, ^n_p}\,
(\delta_{b'\, p}\, \delta_{b\, p}\, -\, \delta_{b'\, n}\, \delta_{b\, n})
\right]
\end{eqnarray}

The second term of the axial-charge operator proportional to
$q_0\vec{\partial_\nu}/(q^2-m_\pi^2)$ contracts with the $\gamma^\nu$
matrix in the $\rho NN$ vertex giving rise to a time-like and a
space-like contribution.  The first one is simply a renormalization of
the matrix element of the operator ${\sf C}_{J}^{A}$ just calculated
and that we now redefine so that,
\begin{equation}
<[a',b']j_2||\hat{\sf C}_{J}^{A*}(q)||[a,b]j_1>=
<[a',b']j_2||\hat{\sf C}_{J}^{A}(q)||[a,b]j_1> \left(
1-\frac{q_0^2}{q^2-m_\pi^2}\right); \label{C(A*)}
\end{equation}
the second (${\sf C}_{J}^{B}$) results in the following expression,
\begin{eqnarray}
& & <[a',b']j_2||\hat{\sf C}_{J}^{B}(q)||[a,b]j_1>=
\frac{i}{4\sqrt{\pi}}g_A g_\rho^2 \frac{q_0\, q}{q^2 - m_\pi^2}
\:\xi_\pm\:  (-1)^J\; \int
dr\:  r^2 \nonumber \\
& & \times \int dr_1\: r_1^2\: M_\pi^{\omega_a} \sum_{l_1=0}^\infty\:
\sqrt{2l_1+1}\: j_{l_1}(iM_\pi^{\omega_a} r^1_<)
h_{l_1}(iM_\pi^{\omega_a} r^1_>)
I_{l_1}(k_{a'},-k_a) \nonumber \\
& &\times \left\{\phantom{\frac{\stackrel{\leftrightarrow}{\partial}}
{\partial r_1}}\!\!\!\! (E_{a'}-E_a)[G_{a'}(r_1)F_a(r_1)-
F_{a'}(r_1)G_a(r_1)]\right. \nonumber \\
& &\;\;\;\;+\left.\left[G_{a'}(r_1)\left(\frac{\stackrel{\leftrightarrow}
{\partial}}{\partial r_1} + \frac{2}{r_1}\right)G_a(r_1) +
F_{a'}(r_1)\left(\frac{\stackrel{\leftrightarrow}{\partial}}
{\partial r_1} + \frac{2}{r_1}\right)F_a(r_1)\right]\right\}
\nonumber \\
& & \times \int dr_2\:  r_2^2\: M_\rho^{\omega_b} \sum_{l_2=0}^\infty
\sqrt{2l_2+1}\: j_{l_2}(iM_\rho^{\omega_b}{r^2}_<)\:
h_{l_2}(iM_\rho^{\omega_b}{r^2}_>)\;
\nonumber \\
& & \times \sum_{l=0}^\infty (2l+1)\:
\frac{1}{\sqrt{(l+l_2+2)(l+l_2+1)(l+l_2)}} \nonumber \\
& & \times \left[G_{b'}(r_2)F_b(r_2)H_{l,l_2}(k_{b'},-k_b)
- F_{b'}(r_2)G_b(r_2) H_{l,l_2}(-k_{b'},k_b)\right]
\nonumber \\
& & \times \sqrt{2} I'_{l,l_2}(k_{b'},-k_b) Z_J(l,l_1,l_2)
\left\{ \begin{array}{ccc}
        j_{a'} & j_{b'} & j_2 \\
        j_a    & j_b    & j_1 \\
        l_1    & l_2    & J   \\
\end{array} \right\} \label{C(B)}
\end{eqnarray}
\noindent
where the functions $H_{l,l_2}(k_{b'},k_b)$, $I'_{l,l_2}(k_{b'},k_b)$ and
$Z_J(l,l_1,l_2)$ are:
\begin{equation}
H_{l,l_2}(k_{b'},k_b)=\delta_{ll_2}\sqrt{2}(k_{b'}-k_b) +
\delta_{l\neq l_2}\left[-k_b'-k_b + (-1)^{\frac{1+l_2-l}{2}}
\left(\frac{l+l_2+1}{2}\right)\right]
\end{equation}
\begin{equation}
I'_{l,l_2}(k_{b'},k_b)=\frac{(-1)^{j_{b'}+1/2}}{4\sqrt{\pi}}\sqrt{(2j_{b'}+1)
(2j_b+1)(2l_2+1)}
\left[1+(-1)^{l_{b'}+l_b+l_2}\right]
\left( \begin{array}{ccc}
       j_{b'}          & l & j_b            \\
       \frac{1}{2} & 0 & -\frac{1}{2} \\
       \end{array} \right)
\end{equation}
\newpage
\begin{eqnarray}
Z_J(l,l_1,l_2) & = & \sqrt{(J+1)(2J+3)} j_{J+1}(qr)
 \left( \begin{array}{ccc}
         l_1 & l_2 & J+1 \\
         0   & 0   & 0 \\
\end{array} \right)
 \left\{ \begin{array}{ccc}
         l_1 & l   & J   \\
         1   & J+1 & l_2 \\
\end{array} \right\} \nonumber \\
& & + \: \sqrt{J(2J-1)} j_{J-1}(qr)
 \left( \begin{array}{ccc}
         l_1 & l_2 & J-1 \\
         0   & 0   & 0 \\
\end{array} \right)
 \left\{ \begin{array}{ccc}
         l_1 & l   & J   \\
         1   & J-1 & l_2 \\
\end{array} \right\}
\end{eqnarray}

Finally, the sum of (\ref{C(A*)}) and (\ref{C(B)}) gives the two-body
matrix element of the axial-charge operator.

\section{Results and discussion}

The coupling constants used in the calculations are shown in
Table~\ref{table1}.  Both sets of parameters, DH (Dirac-Hartree) and
RHA (relativistic Hartree approximation), are from Ref.\cite{horowitz}
and are not tuned specifically to any particular mass region.  The
scalar masses were determined by fitting the charge radius of
$^{40}$Ca.  The other masses, fixed to their experimental values are,
$M=939$ MeV, $m_{\omega}=783$ MeV, $m_{\rho}=770$ MeV and
$m_{\pi}=139$ MeV. In the calculations we used $g_{\rho}^2$ from the
fitting to the bulk symmetry energy $a_4=35$ MeV. The intrinsic
structure of the nucleons was considered via an additional $q^2$
dependence of the form $(1+q^2/M_A^2)^{-2}$ with $M_A=1.2$ GeV.

Tables~\ref{table2} to~\ref{table5} summarize our results.  In the
tables each row displays the result of the calculation of the matrix
element corresponding to the multipole operator shown on the left.
They are the axial Coulomb operator (Eq.\ref{coul}) and
the polar-vector longitudinal
\begin{equation}
{\sf L}_{JM}^{2B}(q)= \int d^{3}x \left[ \frac{i}{q} \vec{\nabla}
 M_J^M(q{\vec{x}}) \right] \cdot {\vec{J}}^{2B}({\vec{x}}),
\label{long}
\end{equation}
and transverse electric,
\begin{equation}
{\sf T}^{2B}_{JM}(q) = \int d^{3}r \left[ \frac{1}{q} \vec{\nabla}
\times {\vec{M}}_J^M(q{\bf r}) \right] \cdot
{\vec{J}}^{2B}({\vec{x}}),
\label{tel}
\end{equation}
operators. In the above we used the expression
\begin{equation}
{\vec{M}}_J^M(q{\vec{x}})=j_L(q|\vec{x}|) {\vec{Y}}_{JL}^M(\hat{x}),
\end{equation}
with
\begin{equation}
{\vec{Y}}_{JL}^M=\sum_{m\mu} C_{m\mu M}^{L1J} Y_L^M(\hat{x})e_\mu.
\end{equation}
the vector spherical harmonics.

Each table is divided in three columns:  The first one (NR) shows the
results of a one-body non-relativistic calculation and has been
included, mainly, for comparative purposes.  The second (DH) shows the
results of:  i) a one-body (IA) calculation, ii) one- plus two-body
(IA+MEC) calculation and iii) the resulting increment ($\delta_A$),
all three obtained within the relativistic Dirac-Hatree approximation
where the vacuum structure is ignored.  The third column has a
structure similar to the previous one but the one-loop vacuum
corrections that originate due to the presence of the strong meson
fields are taken into account~\cite{price}.

The increment $\delta_A$ defined in the tables parametrizes the
difference between the ``effective" axial coupling constant, as
determined from the analysis of the experimental transition rates, and
the ``free" coupling constant such that,
\begin{equation}
g_A^{eff}=g_A^{free}(1+\delta_A).
\end{equation}
We recall that the ``best fit" value reported by Warburton in his
analysis of nuclei in the A=206-212 mass region was
\[ \delta_A=1.05 \pm 0.05.\]

The first thing to notice from the tables is that, as indicated above,
the longitudinal and transverse electric contributions in the one-body
case are substantially smaller than their Coulomb counterparts (more
than a factor of ten in lead and tin; four times smaller in thallium).
Based on this result we restrict the two-body calculation to
the contribution of the axial charge where, on the other hand, we
expect the larger contributions coming from pion exchange.

Despite not being the main thrust of this paper we devote one
paragraph to comment on the one-body vector contribution where we
notice the following.  In the lead transition (table~\ref{table2}),
the IA result for $\hat{L}^{V}_{1}$ and $\hat{T}^{V\ el}_{1}$ is
enhanced a barely 3\% in DH (1\% in RHA) with respect to the NRIA
calculation, a result that is consistent with Warburton's finding for
the polar vector enhancement, namely $\delta_V\approx 0$.  This
behaviour does not maintain, however, and we find that in the thallium
decay to lead the enhancement is 65\% (34\% in RHA) and in tin is 23\%
(18\% in RHA) over the NRIA.  Thus, there is a discrepancy between
this calculation and the fit to the data in the heavy-mass
region~\cite{warburton} for the vector contribution.  {\em A priori}
this difference could be adscribed to the simple-minded model used
here to describe the structure of the states involved; however is hard
to see how a refinement in this direction could reduce drastically the
last two values.

Back to the analysis of the tables the next thing we notice is that, as
expected, the one-body calculation is consistently larger in the
Dirac-Hartree approach than in the relativistic Hartree approximation.
This is due to the value of the effective mass $M^*$ which is smaller in
RHA than in DH.  In principle, there is no theoretical argument favoring
the use of one to the other given that, in both cases, the model
parameters are fitted independently to bulk nuclear-matter properties.
For convenience, in the analysis that follows we shall limit our
discussion to the RHA results.

In the Introduction we argued that in the relativistic framework pair
diagrams are already taken into account in IA; hence the analysis must
proceed by studying the contribution of MEC along with the IA results.
Thus, the $\nu:2g_{9/2}\rightarrow \pi:1h_{9/2}$ transition in
$^{209}{\rm Pb}\rightarrow$ $^{209}{\rm Bi}$ shows that some 40\% of
the enhancement is present at the one-body level and a further 80\%
shows up at the two-body level bringing their sum up to a total
enhancement of $\delta_A=1.2$.  The other transition we studied in a
heavy nucleus, $\pi:(3s_{1/2})^{-1} \rightarrow \nu:(3p_{1/2})^{-1}$
in $^{207}{\rm Tl}\rightarrow$ $^{207}{\rm Pb}$ gives a 30\%
enhancement due to one-body processes and 55\% coming from the meson
exchange contributions.

If we move to lighter nuclei the situation does not change much as far
as the overall increment of the effective coupling constant is
concerned.  In the $^{133}{\rm Sn} \rightarrow ^{133}{\rm Sb}$ decay,
dominated by the single particle transition $\nu:2f_{7/2} \rightarrow
\pi:1g_{7/2}$, we obtained 42\% enhancement from IA and 33\% from the
MEC processes.  In the oxygen region the decay $^{16}{\rm N}(0^-)
\rightarrow ^{16}{\rm O}(0^+)$ was studied using a simple model of
nuclear structure where the first excited state of $^{16}{\rm N}$,
$J^\pi=0^-$ (with an excitation energy of 120 keV) was described by
the mixing of two particle-hole configurations,
\begin{eqnarray}
|\, 0^-\, \rangle &=&
\sqrt{1 - \lambda^2}\,\, |[2s_{1/2} 1p_{1/2}]_{0^-}\, \rangle\,
+\, \lambda\,\, |[1d_{3/2} 1p_{3/2}]_{0^-}\, \rangle
\label{lambda}
\end{eqnarray}

\noindent
and where the mixing amplitude $\lambda$ was taken to be the typical
$\lambda\approx 0.08$~\cite{TownerKhanna}. In this case we obtained
48\% (IA) and 42\% (MEC) for the largest component, and 18\% (IA) and
10\% (MEC) for the smallest.

Overall the calculated value of $\delta_A$ is around 20\% of that
determined empirically in the heavy-mass region.  The ratio of one- to
two-body processes, however, is not distributed uniformly in mass.  As
one moves towards the heavier nuclei the share between the one-body
and two-body processes is shifted from the former to the latter.  It
is not straightforward to explain this trade off.  Naively, it could
be argued that the increase in the density implies a reduced effective
mass and a corresponding enhancement in the nucleon propagators; one
such change, however, should affect both the one-body and two-body
processes in roughly the same manner.  Furthermore, the argument of
enhanced matrix elements due to off-diagonal operators (operators
connecting upper to lower components of the spinors) it is not
strictly true for $\beta$-decay transitions where the initial and
final states do not coincide.  The value of the matrix element depends
on the overlap of initial and final wave functions and this overlap
diminishes as we go up in mass in the periodic table.  A case in point
are the different enhancements obtained for the vector operators,
$\hat{L}^{V}_{1}$ and $\hat{T}^{V\ el}_{1}$, which are very different
in different transitions despite being both off-diagonal.
Summarizing, the results of the tables indicate a somewhat stable
contribution of the one-body processes of around 40\% whereas the
two-body share seems to increase slightly with the mass (O:42\%;
Sn:33\%; Tl:55\%; Bi:80\%).

The calculation in oxygen with a simple nuclear model gives an
enhancement close to 100\%.  Before the work of Haxton and
Johnson~\cite{haxton} in 1990, this result would have been considered
anomalous.  These authors found that the usually accepted 50\%
enhancement of the coupling constant fell short by some 40\% of the
value necessary to agree with their calculation of the $\beta$-decay
rate of the 120 keV 0$^-$ state in nitrogen.  This discrepancy plus
the one found in heavy nuclei prompted a more careful examination of
previous calculations.  Presently, this discrepancy is not considered
dramatic in any mass region~\cite{warbtow} though there seems to be
still some room for other effects.  Our results seem to indicate a
similar behaviour (so long as we stick to the RHA model and use
pseudovector coupling).

Regarding the type of coupling, Table~\ref{table6} illustrates the
difference between pseudoscalar and pseudovector contributions for a
particular transition in lead.  Use of the former leads, typically, to
an increment of the calculated enhancement of approximately 50\% with
respect to the same calculation with pseudovector coupling.  This is
not, {\em per se}, ground enough to preclude the use of pseudoscalar
coupling.  However, for the many-body problem is more
efficient~\cite{serot2} to work with a Lagrangian with pseudovector
coupling and no$\sigma\pi\pi$ term because in this form the soft-pion
limits appear naturally in terms of just pions and nucleons (without
the necessity of subtle cancellations with terms containing the
$\sigma$).

One point to notice is the dependence of the MEC results on the value
of $g_\rho^2$.  Calculations in finite nuclei within QHD (see, for
example, Refs.\cite{serot} and \cite{price}) usually employ
$g_\rho^2=65.23$ (the value used in this work) instead of
$g_\rho^2=36.76$ as calculated from the decay of the $\rho$ meson.
The first value is obtained fitting the symmetry energy of the
empirical mass formula; this is a many-body argument which supports
its use over that obtained from $\rho$ decay.  However, aside from
this, there are no further arguments which are strong enough to
discriminate them; this ambiguity translates into a downwards 40\%
uncertainty in the final MEC results.  It is relevant to comment that the
relativistic one-body calculation is not sensitive to the value of the $\rho$
coupling constant.

A final word about core polarization.  In two previous
works~\cite{izquierdo,izquierdo1} we investigated the {\em isovector}
weak response of the core to the external perturbation in the
approximation of nuclear matter and in linear response theory summing
the ring diagrams to all orders in the random phase approximation
(RPA) ---the so-called {\em backflow}
correction~\cite{furnstahl,cohen}.  We found that the effects of this
core polarization (of positive and negative energy nucleons) were
sizable for large values of the momentum transfer (for example those
typical of the muon capture processes, $q^2\approx 0.9 m_\mu^2$) but
could be neglected for the small momentum transfers involved in
nuclear $\beta$ decay.  Thus, we do not anticipate any dramatic
increase in the calculated $\delta_A$ beyond those presented in this
work.

Summarizing, we have studied the contribution of two-body (MEC)
processes to axial charge transitions for nuclei in the lead, tin and
oxygen regions.  We conducted calculations in Dirac-Hartree
approximation (the Walecka model) and in the relativistic Hartree
approximation (where the full one-nucleon-loop effects are included.)
Along the paper we discussed the sensitivities of the calculation to
the change in some of the couplings and of the parameters involved.
With the choices described above the results we obtained indicate that
one- and two-body processes enhance the matrix elements of the
axial-charge operator by some (100$\pm$20) \% in all three regions
studied.  This result agrees with the enhancement of the
axial-coupling constant needed to understand theoretically a fit to
eighteen first-forbidden $\beta$ decay transitions in the lead region.
{}From the results of this calculation the presence of a heavy scalar
meson to explain a physical effect gets further support.

\acknowledgements

Partial financial support from the Antorchas Foundation is gratefully
acknowledged.  A.O.G. is fellow of the CONICET, Argentina.

\begin{figure}
\caption { a) and b): Pair graphs representing the exchange of a
scalar meson.  Diagrams c) and d) are those evaluated in
relativistic mean-field theories (one body).}
\label{fig1}
\end{figure}

\begin{figure}
\caption { A ``time-ordered" diagram involving an intermediate
nucleon-antinucleon pair generated by the presence of the scalar
field. This diagram is
included in the relativistic meson-exchange calculation.}
\label{fig2}
\end{figure}
\begin{figure}
\caption { Tree-level diagrams to be considered in the calculation of
two-body processes:
a) $\rho\pi$, b) $\rho\pi\pi$, c) $\Delta\pi$, and d) $\Delta\pi\pi$.}
\label{fig3}
\end{figure}

\begin{figure}
\caption{ a) $\pi NN$, b) $\rho NN$ and c) $\rho\pi\pi$ vertices
derived from the Lagrangian of QHD-II [Eq.~(\protect\ref{Lagrangian}) in
the text].}
\label{fig4}
\end{figure}
\begin{figure}
\caption { The weak $\pi\varepsilon$ and $\pi\rho\varepsilon$ vertices
needed for the calculation and not present explicitly in the QHD-II
Lagrangian.}
\label{fig5}
\end{figure}
\narrowtext
\begin{table}
\caption{Parameters used for the calculations presented in
the text.}
\begin{tabular}{lllllll}
& $g_{\sigma N}^2$ & $m_\sigma$ & $g_{\omega N}^2$ & $g_\pi^2$
& $g_\rho^2$ & $M^*/M$
\\
\tableline
MFT & 109.6 & 520 & 190.4 & 178.4 & 65.23 & 0.54 \\
RHA &  54.3 & 458 & 102.8 & 178.4 & 65.23 & 0.73 \\
\end{tabular}
\label{table1}
\end{table}

\begin{table}
\caption[]{Results for the $\nu:2g_{9/2}\rightarrow \pi:1h_{9/2}$
transition in $^{209}{\rm Pb}\rightarrow$ $^{209}{\rm Bi}$. First
row: results for the axial-charge operator. All the impulse
approximation (IA) results, non-relativistic (NR), Dirac-Hartree (DH)
and relativistic Hartree (RHA) are from
reference~\protect\cite{izquierdo}. The contribution of pion-exchange
currents is summed in the columns labeled $IA+MEC$. The total
increment $\delta_A$ is given for the two relativistic calculations.
Second and third rows: same as above but for the longitudinal
($\hat{L}_1^V$) and transverse electric ($\hat{T}_1^{V\;el}$)
operators.}
\vspace{.5cm}
\begin{tabular}{c|c|ccc|ccc}
&\multicolumn{1}{c|}{${NR}$} & \multicolumn{3}{c|}{${DH}$} &
\multicolumn{3}{c}{${RHA}$} \\
\tableline
& $\scriptstyle{IA}$ & $\scriptstyle{IA}$ & $\scriptstyle{IA + MEC}$ &
$\scriptstyle{\delta_A}$ &
$\scriptstyle{IA}$ & $\scriptstyle{IA + MEC}$ & $\scriptstyle{\delta_A}$ \\
$\hat{C}^{A}_{0}$ & $\scriptstyle{\,\,1.12\times10^{-1}}$ &
$\scriptstyle{\,\,1.98\times10^{-1}\,\, }$ &
$\scriptstyle{\,\,3.12\times10^{-1}}$ & $\scriptstyle{1.77}\,\,\,\,$ &
$\scriptstyle{\,\,1.58\times10^{-1}\,\, }$ &
$\scriptstyle{\,\,2.48 \times10^{-1}}$ & $\scriptstyle{1.2\,\,\,\,}$ \\
$\hat{L}^{V}_{1}$ &
$\scriptstyle{\,\,5.61\times10^{-3}}$ &
$\scriptstyle{\,\,5.77\times10^{-3}\,\, }$ & & &
$\scriptstyle{\,\,5.69\times10^{-3}\,\, }$ & &\\
$\hat{T}^{V\, el}_{1}$ &
$\scriptstyle{\,\,7.93\times10^{-3}}$ &
$\scriptstyle{\,\,8.16\times10^{-3}\,\, }$ & & &
$\scriptstyle{\,\,8.05\times10^{-3}\,\, }$ & &\\
\end{tabular}
\label{table2}
\end{table}

\begin{table}
\caption{Same as Table~\protect\ref{table2} for the transition
$^{207}{\rm Tl}(1/2^+)$ $\rightarrow$ $^{209}{\rm Pb}(1/2^-)$}
\vspace{.5cm}
\begin{tabular}{c|c|ccc|ccc}
&\multicolumn{1}{c|}{${NR}$} & \multicolumn{3}{c|}{${DH}$} &
\multicolumn{3}{c}{${RHA}$} \\
\tableline
& $\scriptstyle{IA}$ & $\scriptstyle{IA}$ & $\scriptstyle{IA + MEC}$ &
$\scriptstyle{\delta_A}$ &
$\scriptstyle{IA}$ & $\scriptstyle{IA + MEC}$ & $\scriptstyle{\delta_A}$ \\
$\hat{C}^{A}_{0}$ & $\scriptstyle{-0.75\times10^{-1}}$ &
$\scriptstyle{-1.15\times10^{-1}\,\,}$ &
$\scriptstyle{-1.60\times10^{-1}}$ & $\scriptstyle{1.12}\,\,\,\,$ &
$\scriptstyle{-0.97\times10^{-1}\,\, }$ &
$\scriptstyle{-1.37 \times10^{-1}}$ & $\scriptstyle{0.81\,\,\,\,}$ \\
$\hat{L}^{V}_{1}$ & $\scriptstyle{2.01\times10^{-2}}$ &
$\scriptstyle{3.33\times10^{-2}\,\, }$ & & &
$\scriptstyle{2.71\times10^{-2}\,\, }$ & &\\
$\hat{T}^{V\, el}_{1}$ & $\scriptstyle{2.85\times10^{-2}}$ &
$\scriptstyle{4.71\times10^{-2}\,\, }$ & & &
$\scriptstyle{3.83\times10^{-2}\,\, }$ & &\\
\end{tabular}
\label{table3}
\end{table}

\begin{table}
\caption{Same as Table~\protect\ref{table2} for the transition
$^{133}{\rm Sn}(7/2^-)$ $\rightarrow$ $^{133}{\rm Sb}(7/2^+)$}
\vspace{.5cm}
\begin{tabular}{c|c|c c c|c c c}
&\multicolumn{1}{c|}{${NR}$} & \multicolumn{3}{c|}{${DH}$} &
\multicolumn{3}{c}{${RHA}$} \\
\tableline
& $\scriptstyle{IA}$ & $\scriptstyle{IA}$ & $\scriptstyle{IA + MEC}$ &
$\scriptstyle{\delta_A}$ &
$\scriptstyle{IA}$ & $\scriptstyle{IA + MEC}$ & $\scriptstyle{\delta_A}$ \\
$\hat{C}^{A}_{0}$ &
$\scriptstyle{\,\,0.98\times10^{-1}}$ &
$\scriptstyle{\,\,1.69\times10^{-1}\,\, }$ &
$\scriptstyle{\,\,2.22\times10^{-1}}$ & $\scriptstyle{1.26}\,\,\,\,$ &
$\scriptstyle{\,\,1.42\times10^{-1}\,\, }$ &
$\scriptstyle{\,\,1.70\times10^{-1}}$ & $\scriptstyle{0.73}\,\,\,\,$ \\
$\hat{L}^{V}_{1}$ &
$\scriptstyle{\,\,0.66\times10^{-2}}$ &
$\scriptstyle{\,\,0.81\times10^{-2}\,\, }$ & & &
$\scriptstyle{\,\,0.78\times10^{-2}\,\, }$ & &\\
$\hat{T}^{V\, el}_{1}$ &
$\scriptstyle{\,\,0.93\times10^{-2}}$ &
$\scriptstyle{\,\,1.15\times10^{-2}\,\, }$ & & &
$\scriptstyle{\,\,1.10\times10^{-2}\,\, }$ & &\\
\end{tabular}
\label{table4}
\end{table}

\begin{table}
\caption{Same as Table~\protect\ref{table2} for the transition
$^{16}{\rm N}(0^-)$ $\rightarrow$ $^{16}{\rm O}(0^+)$ except that the
one-body longitudinal ($\hat{L}^{V}_{1}$) and transverse electric
($\hat{T}^{V\, el}_{1}$) contributions have been substituted by the
$J=3/2$ contribution from the transition due to the second term in
Eq.~(\protect\ref{lambda})}
\vspace{.5cm}
\begin{tabular}{c|c|c c c|c c c}
&\multicolumn{1}{c|}{${NR}$} & \multicolumn{3}{c|}{${DH}$} &
\multicolumn{3}{c}{${RHA}$} \\
\tableline
& $\scriptstyle{IA}$ & $\scriptstyle{IA}$ & $\scriptstyle{IA + MEC}$ &
$\scriptstyle{\delta_A}$ &
$\scriptstyle{IA}$ & $\scriptstyle{IA + MEC}$ & $\scriptstyle{\delta_A}$ \\
$\hat{C}^{A}_{0}\, \scriptstyle{j = 1/2}$ &
$\scriptstyle{\,\,3.95\times10^{-2}}$ &
$\scriptstyle{\,\,5.97\times10^{-2}\,\, }$ &
$\scriptstyle{\,\,7.90\times10^{-2}}$ & $\scriptstyle{1.35}\,\,\,\,$ &
$\scriptstyle{\,\,5.86\times10^{-2}\,\, }$ &
$\scriptstyle{\,\,7.51\times10^{-2}}$ & $\scriptstyle{0.90}\,\,\,\,$ \\
$\hat{C}^{A}_{0}\, \scriptstyle{j = 3/2}$ &
$\scriptstyle{- 1.31\times10^{-1}}$ &
$\scriptstyle{- 1.64\times10^{-1}\,\, }$ &
$\scriptstyle{- 1.78\times10^{-1}}$ & $\scriptstyle{0.36}\,\,\,\,$ &
$\scriptstyle{- 1.56\times10^{-1}\,\, }$ &
$\scriptstyle{- 1.68\times10^{-1}}$ & $\scriptstyle{0.28}\,\,\,\,$ \\
\end{tabular}
\label{table5}
\end{table}

\narrowtext
\begin{table}
\caption[]{Comparison of the MEC results using pseudovector (PV)
coupling and pseudoscalar plus the $\sigma\pi\pi$ term for the axial
charge operator.  The results correspond to the
$\nu:2g_{9/2}\rightarrow \pi:1h_{9/2}$ transition in $^{209}{\rm
Pb}\rightarrow$ $^{209}{\rm Bi}$ and the labels are the same as in
Table~\protect\ref{table2}. The second row shows the enhancement
with respect to the $NR$ calculation.}
\begin{tabular}{ccccccc}
\multicolumn{1}{c}{$NR$}  & \multicolumn{3}{|c}{$DH$} &
\multicolumn{3}{|c}{$RHA$}  \\
\tableline
\multicolumn{1}{c}{$IA$} & \multicolumn{1}{c}{$IA$}     &
\multicolumn{1}{c}{$MEC(PV)$} & \multicolumn{1}{c}{$MEC(PS)$} &
\multicolumn{1}{c}{$IA$}     & \multicolumn{1}{c}{$MEC(PV)$} &
\multicolumn{1}{c}{$MEC(PS)$} \\
\tableline
-1.23$\times 10^{-1}$ & -1.98$\times 10^{-1}$ & -1.14$\times 10^{-1}$ &
-2.63$\times 10^{-1}$ & -1.57$\times 10^{-1}$ & -8.97$\times 10^{-2}$ &
-2.14$\times 10^{-1}$  \\
& 0.60 & 0.90 & 2.14 & 0.28 & 0.72 & 1.72 \\
\end{tabular}
\label{table6}
\end{table}


%
%
%
%
%
%
%
%
%
%


\begin{references}
\bibitem{warburton} E.\ K.\ Warburton, Phys. Rev. Lett. {\bf 66}, 1823
(1991); Phys. Rev. {\bf C44}, 233 (1991).
\bibitem{meyer}H.\ O.\ Meyer, C.\ Horowitz, H.\ Nann, P.\ V.\ Pancella,
S.\ F.\ Pate, R.\ E. Pollock, B.\ Von Przewoski, T.\ Rinckel, M.\ A.\
Ross and F.\ Sperisen, Nucl. Phys. {\bf A539}, 633 (1992).
\bibitem{towner1} I.S. Towner, Comments Nucl. Part. Phys. {\bf 15}, 145
(1986).
\bibitem{riska} M.\ Kirchbach, D.\ O.\ Riska and K.\ Tsushima, Nucl.
Phys. {\bf A542}, 616 (1992).
\bibitem{towner} I.\ S.\ Towner, Nucl. Phys. {\bf A542}, 631 (1992).
\bibitem{chuck}C.\ Horowitz, H.\ O.\ Meyer and D.\ K.\ Griegel, Phys. Rev.
{\bf C48}, 2920 (1993).
\bibitem{delTow} J. Delorme and I.S. Towner, Nucl. Phys. {\bf A475}, 720
(1987).
\bibitem{arima}W.\ Bentz, A. Arima, H.\ Hyuga, k.\ Shimizu and k.\
Yazaki, Nucl. Phys. {\bf A436}, 593 (1985); S.\ Ichii, W. Bentz and
A.\ Arima, Nucl. Phys. {\bf A464} 575 (1987).
\bibitem{blunden}P. Blunden, Nucl. Phys. {\bf A464}, 525 (1987).
\bibitem{warbtow}E.\ K.\ Warburton and I.\ S.\ Towner, Phys. Reports
{\bf 243}, 103 (1994).
\bibitem{serot2} B.\ D.\ Serot, Rep. Prog. Phys., {\bf 11} 11855 (1992).
\bibitem{izquierdo} A.\ O.\ Gattone, E.\ D.\ Izquierdo and M.\
Chiapparini, Phys. Rev. {\bf C46}, 788 (1992).
\bibitem{izquierdo1} E.\ D.\ Izquierdo and A.\ O.\ Gattone, Phys. Rev.
{\bf C49}, 2005 (1994).
\bibitem{barenboim} G.\ Barenboim, A.\ O.\ Gattone and E.\ D.\
Izquierdo, Phys. Rev. {\bf C48}, 2537 (1993).
\bibitem{nozawa} S.\ Nozawa, K.\ Kubodera and H.\ Ohtsubo, Nucl.  Phys.
{\bf A453}, 645 (1986).
\bibitem{serot} B.\ D.\ Serot and J.\ D.\ Walecka, in {\em The
Relativistic Nuclear Many--Body Problem}, Advances in Nuclear Physics,
Vol. 16.  J.W.  Negele and E. Vogt eds., Plenum Press, New York 1986.
\bibitem{furnstahl}R.J. Furnstahl and C.E. Price, Phys. Rev {\bf C41},
1792 ( 1990).
\bibitem{Walecka Muon} J.D. Walecka, in {\em Muon Physics}, Vol. II,
V.W. Hughes and C.S. Wu eds. Academic Press, New York 1975, p. 113.
\bibitem{ivanov} Ivanov and Truhlik, Nucl. Phys. {\bf B234} 167 (1975).
\bibitem{donnelly}T.\ W.\ Donnelly and I.\ Sick, Rev. Mod. Phys. {\bf
56}, 461 (1984).
\bibitem{horowitz} C.\ J.\ Horowitz and B.\ D.\ Serot, Phys. Lett. {\bf
140B}, 181 (1984).
\bibitem{price} R.\ J.\ Furnstahl and C.\ E.\ Price, Phys. Rev. {\bf
C41}, 1792 (1990). The same authors use this value to
calculate density distributions in, Phys. Rev. {\bf C44}, 895 (1991), and a
slightly larger one ($g_\rho^2=83.30$) in Phys. Rev. {\bf C40}, 1398
(1989).
\bibitem{TownerKhanna}I.\ S.\ Towner and F.\ C.\ Khanna, Nucl. Phys.
{\bf A372}, 331 (1981).
\bibitem{haxton}W.\ C.\ Haxton and C.\ Johnson, Phys. Rev. Lett. {\bf
65}, 1325 (1990).
\bibitem{furnstahl} R.\ J.\ Furnstahl and B.\ D.\ Serot, Nucl. Phys.
{\bf A468}, 539 (1987).
\bibitem{cohen} J.\ Cohen, Phys. Rev. {\bf C48}, 1346 (1993).

\end{references}
\end{document}